\documentclass{article}

\usepackage{arxiv}

\usepackage[utf8]{inputenc}
\usepackage[T1]{fontenc}
\usepackage{hyperref}
\usepackage{url}
\usepackage{booktabs}
\usepackage{amsmath,amssymb}
\usepackage{amsfonts}
\usepackage{graphicx}
\usepackage{natbib}
\usepackage{doi}
\usepackage{cleveref}
\usepackage{microtype}
\usepackage{nicefrac}

\title{Spectral-Domain Local Statistics with Missing-Data Support\\
for Cartesian and Polar Grids}

\author{
  \setlength{\tabcolsep}{3pt}%
  \begin{tabular}[t]{ccc}
    Jairo M.~Valdivia-Prado$^{1}$\,\href{https://orcid.org/0000-0003-0709-1163}{\includegraphics[height=8pt]{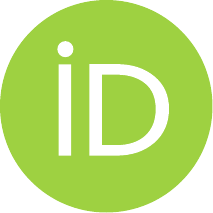}} &
    William E.~Chapman$^{1}$\,\href{https://orcid.org/0000-0002-0472-7069}{\includegraphics[height=8pt]{orcid.pdf}} &
    Katja Friedrich$^{1}$\,\href{https://orcid.org/0000-0001-8417-4232}{\includegraphics[height=8pt]{orcid.pdf}} \\[0.15cm]
    \small\normalfont\texttt{jairo.valdiviaprado@colorado.edu} &
    \small\normalfont\texttt{william.chapman@colorado.edu} &
    \small\normalfont\texttt{katja.friedrich@colorado.edu}
  \end{tabular} \\[0.3cm]
  \normalfont $^{1}$Department of Atmospheric and Oceanic Sciences, University of Colorado Boulder, Boulder, CO, USA
}

\renewcommand{\shorttitle}{\textit{Spectral-Domain Local Statistics}}

\hypersetup{
  colorlinks=true,
  linkcolor=blue,
  citecolor=black,
  urlcolor=blue,
  pdftitle={Spectral-Domain Local Statistics with Missing-Data Support for Cartesian and Polar Grids},
  pdfauthor={Jairo M.~Valdivia-Prado, William E.~Chapman, Katja Friedrich},
  pdfkeywords={DCT, smoothing, normalized convolution, boundary conditions, local statistics, missing data},
}

\begin{document}
\maketitle

\begin{abstract}
	This paper presents a method for computing local mean, variance, standard deviation, and effective sample count on incomplete gridded data using boundary-aware spectral operators. The framework combines normalized convolution with explicit boundary-condition modeling: reflective Discrete Cosine Transform (DCT) for non-periodic Cartesian axes and periodic Real Fast Fourier Transform (RFFT) for circular azimuth processing in polar geometry. Stability safeguards (denominator floor, prefill fallback, and variance clamp) are specified for under-supported regions. We evaluate the framework across three targeted scenarios: a Cartesian boundary-condition check demonstrating the mitigation of wrap-around artifacts, a synthetic 3D outlier-identification test, and a real-radar polar application. Results establish bounded, support-aware interpretation of local statistics while preserving a concise reproducibility path through the open-source \texttt{dct\_toolkit} implementation.
\end{abstract}


\begin{figure}[!ht]
	\centering
	\includegraphics[width=0.98\linewidth]{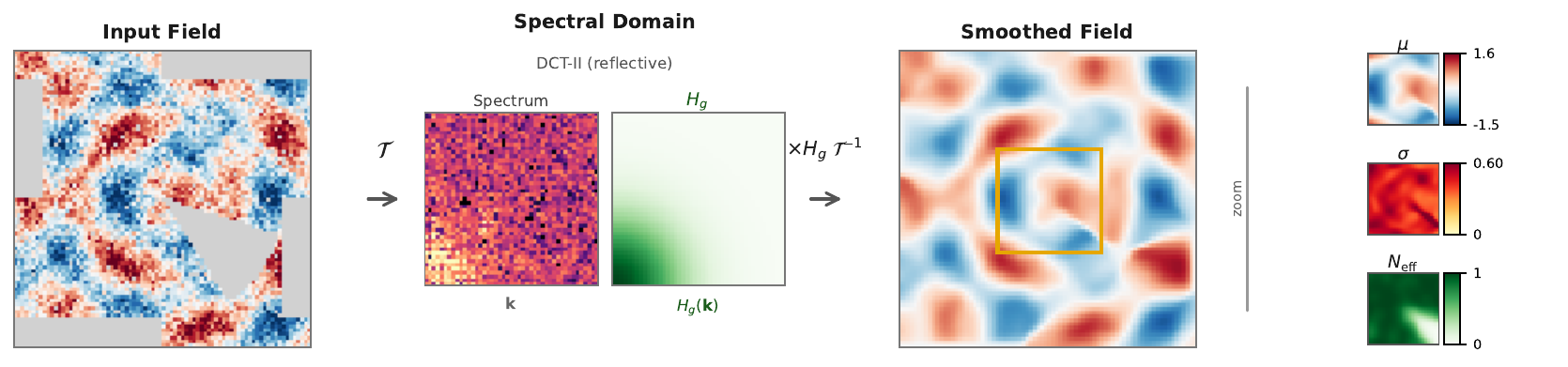}
	\caption{Spectral normalized-convolution pipeline illustrated on a 2-D synthetic field with missing values (gray). The masked field is transformed with DCT-II (reflective boundaries), multiplied by a Gaussian transfer function $H_g$, and inverted to yield the smoothed field. The zoomed inset shows the corresponding local mean $\mu$, standard deviation $\sigma$, and effective sample count $N_{\mathrm{eff}}$.}
	\label{fig:teaser}
\end{figure}

\clearpage

\section{Introduction}

Large gridded datasets in meteorology, radar remote sensing, Earth observation, and related scientific imaging workflows routinely contain missing or invalid values due to cloud contamination, quality-control masking, beam blockage, clutter filtering, and acquisition geometry \citep{Asner2001,JuRoy2008,Shen2015,Steiner2002,Hubbert2009Part1,Bech2003}. In these workflows, local mean, variance, and standard deviation are foundational diagnostics for smoothing and quality control \citep{Savitzky1964,Harris1978,Bringi2001}.

For large support windows, spatial convolution scales as $\mathcal{O}(N\cdot K)$, while spectral convolution scales as $\mathcal{O}(N\log N)$ \citep{Cooley1965,Oppenheim1999,Nussbaumer1982}. However, FFT-based spectral pipelines assume periodic extension by default and can produce edge wrap-around artifacts in non-periodic domains unless explicit boundary treatment is introduced \citep{AlmeidaFigueiredo2013,Vio2005}. Missing-data handling in spectral form also requires explicit confidence weighting to avoid conflating missing values with zeros.

Normalized convolution addresses this by filtering both data and certainty and then normalizing by local support \citep{Knutsson1993,Westin1994,Pham2006}. In practical scientific pipelines, this appears as mask-weighted convolution and support normalization \citep{Astropy2022}. For non-periodic domains, DCT formulations are a natural alternative to FFT formulations because they retain spectral efficiency while matching reflective/Neumann-compatible boundary models \citep{Ahmed1974,Makhoul1980,Strang1999,Ng1999,Martucci1994}.

Many scientific products are analyzed on Cartesian $N$-dimensional grids, while many instruments, especially weather radar, are sampled natively in polar geometry where azimuth and range axes have different physical and boundary behavior \citep{Doviak1979,Bringi2001,Fabry2015}. The prior DCT radar method in \citet{Valdivia2026RadarDCT} established efficient polar smoothing for radar-focused workflows but focused on mean-only outputs and interpolation-first NaN handling. The present paper generalizes that baseline by (i) unifying Cartesian and polar operator forms under explicit boundary modes, (ii) deriving normalized local statistics beyond the mean, (iii) formalizing stability and support diagnostics, and (iv) adding threshold-based 3D outlier-identification validation, expanding applicability beyond radar-only use cases.
Figure~\ref{fig:teaser} previews the spectral normalized-convolution pipeline and the local statistics ($\mu$, $\sigma$, $N_{\mathrm{eff}}$) produced by the framework.

The manuscript is organized as follows. Section~\ref{sec:notation} defines notation and operators. Section~\ref{sec:convolution_theorem} states convolution-theorem mechanics and transfer functions for reflective and periodic modes. Section~\ref{sec:nc_statistics} derives normalized local statistics and stability safeguards. Section~\ref{sec:polar} formulates the polar extension with adaptive azimuth kernels. Section~\ref{sec:validation} reports targeted validation evidence. Section~\ref{sec:discussion} discusses interpretation and limitations, and Section~\ref{sec:conclusion} concludes.

\section{Definitions and Notation}
\label{sec:notation}

Let $x$ denote the observed field and $m \in \{0,1\}$ denote a certainty mask, where $m(\mathbf{i})=1$ indicates valid observations and $m(\mathbf{i})=0$ indicates missing or invalid entries at index $\mathbf{i}$. Define the spectral smoothing operator with kernel $g$ as
\begin{equation}
	\mathcal{S}_g\{f\} = \mathcal{T}^{-1}\!\left\{\mathcal{T}\{f\} \cdot H_g\right\},
	\label{eq:smooth_operator}
\end{equation}
where $\mathcal{T}$ and $\mathcal{T}^{-1}$ are matched forward/inverse transforms and $H_g$ is the transfer function. Under reflective boundaries, $\mathcal{T}$ is DCT-II (ortho normalization); under periodic boundaries, $\mathcal{T}$ is Real Fast Fourier Transform (RFFT, ortho normalization). Figure~\ref{fig:teaser} illustrates this pipeline for a 2-D Cartesian example with missing data and reflective boundaries.

\textbf{Remark on operator usage.} $\mathcal{S}_g$ is a linear spectral filter that can be applied to any field $f$---including the observed data $x$, the certainty mask $m$, squared data $x^2$, or a prefilled field $\bar{x}$. It is not, by itself, a local mean estimate. In the presence of missing data, an unbiased local mean requires a support-normalized combination of two $\mathcal{S}_g$ evaluations (Section~\ref{sec:nc_statistics}).

This section defines operators and assumptions only; performance evidence is deferred to Section~\ref{sec:validation}.

In this manuscript, $g$ denotes one of three kernels: Gaussian, boxcar, or discrete boxcar. Gaussian and boxcar kernels are retained because they provided the strongest efficiency-performance trade-off in the prior radar-focused DCT study \citep{Valdivia2026RadarDCT}. The discrete boxcar kernel is included because, for complete (no-missing) inputs, it matches the standard spatial-domain discrete convolution operation.

Table~\ref{tab:notation} summarizes notation used throughout the manuscript.

\begin{table}[htbp]
	\centering
	\caption{Summary of notation.}
	\label{tab:notation}
	\begin{tabular}{ll}
		\toprule
		Symbol                   & Meaning                                                    \\
		\midrule
		$x$                      & observed data field                                        \\
		$m \in \{0,1\}$          & certainty mask ($1$ valid, $0$ missing)                    \\
		$g$                      & smoothing kernel (boxcar, boxcar\_discrete, gaussian)      \\
		$H_g$                    & spectral transfer function of kernel $g$                   \\
		$\mathcal{S}_g\{\cdot\}$ & spectral smoothing operator (Eq.~\ref{eq:smooth_operator}) \\
		$\mu$                    & normalized local mean estimate                             \\
		$\sigma^2,\sigma$        & local variance and standard deviation                      \\
		$N_{\mathrm{eff}}$       & effective local sample count                               \\
		$\epsilon$               & denominator floor (default $10^{-3}$)                      \\
		\bottomrule
	\end{tabular}
\end{table}

The distinction between the linear smoothing operator $\mathcal{S}_g$ and the normalized local statistics ($\mu$, $\sigma^2$, $N_{\mathrm{eff}}$) defined in Section~\ref{sec:nc_statistics} is central to the missing-data treatment below.

\section{Convolution Theorem and Spectral Smoothing}
\label{sec:convolution_theorem}

The role of this section is to connect the abstract smoothing operator $\mathcal{S}_g$ defined in Section~\ref{sec:notation} to an efficient spectral implementation. In the present framework, local statistics are obtained by multiplying spectral coefficients of the data and the mask by a kernel transfer function $H_g$ and inverting the transform. The choice of transform is not only a computational device but also a boundary-condition model: Discrete Cosine Transform (DCT-II) enforces reflective (half-sample symmetric) extension on non-periodic axes, while Real Fast Fourier Transform (RFFT) enforces periodic extension on circular axes such as azimuth in polar geometry.

The central identity is the convolution theorem: convolution in the spatial domain is equivalent to point-wise multiplication in the spectral domain \citep{Oppenheim1999}. For signals $f[n]$ and $g[n]$,
\begin{equation}
	\mathcal{T}\{f * g\} = \mathcal{T}\{f\} \cdot \mathcal{T}\{g\},
	\label{eq:conv_theorem}
\end{equation}
with $*$ denoting convolution. Complexity context follows standard FFT references \citep{Cooley1965,Nussbaumer1982}.

\subsection{Reflective and Periodic Boundary Interpretation}

When $\mathcal{T}$ is DCT-II with orthonormal normalization, Eq.~\ref{eq:conv_theorem} corresponds to half-sample symmetric (reflective) extension and Neumann-compatible boundary interpretation \citep{Strang1999,Ng1999,Martucci1994}. When $\mathcal{T}$ is DFT/RFFT, the same identity corresponds to periodic extension.

\subsection{Transfer Functions}

\paragraph{Reflective (DCT) transfer functions.}
Three kernel forms are used:
\begin{align}
	H_{\mathrm{box}}[k]   & = \frac{\sin(W\,\theta_k)}{W\sin\theta_k}, \qquad \theta_k = \frac{\pi k}{2N}, \label{eq:tf_boxcar_analytical}                                                     \\
	H_{\mathrm{dis}}[k]   & = \frac{1}{w_{\mathrm{int}}}\left(1+2\sum_{j=1}^{(w_{\mathrm{int}}-1)/2}\cos(j\,\omega_k)\right), \qquad \omega_k = \frac{\pi k}{N}, \label{eq:tf_boxcar_discrete} \\
	H_{\mathrm{gauss}}[k] & = \exp\!\left(-\frac{1}{2}(\omega_k\sigma)^2\right), \qquad \omega_k = \frac{\pi k}{N}, \quad \sigma = \frac{W}{\sqrt{12}}. \label{eq:tf_gaussian}
\end{align}
For the analytical boxcar, $k=0$ gives $\theta_0=0$; the limit yields $H_{\mathrm{box}}[0]=1$, preserving the DC component. The same limit holds for the periodic analytical boxcar in Eq.~\ref{eq:tf_boxcar_periodic}.
Here $w_{\mathrm{int}}$ is an odd integer width, so $(w_{\mathrm{int}}-1)/2$ is integer-valued in Eq.~\ref{eq:tf_boxcar_discrete}.
The Gaussian width mapping follows the boxcar-variance match \citep{Smith1997}.

\paragraph{Periodic (RFFT) transfer functions.}
For periodic azimuth processing,
\begin{align}
	H_{\mathrm{box}}^{\mathrm{per}}[k]   & = \frac{\sin(W\,\theta_k)}{W\sin\theta_k}, \qquad \theta_k = \frac{\pi k}{N_{\mathrm{az}}}, \label{eq:tf_boxcar_periodic}                                                  \\
	H_{\mathrm{gauss}}^{\mathrm{per}}[k] & = \exp\!\left(-\frac{1}{2}(\omega_k\sigma)^2\right), \qquad \omega_k = \frac{2\pi k}{N_{\mathrm{az}}}, \quad \sigma = \frac{W}{\sqrt{12}}. \label{eq:tf_gaussian_periodic}
\end{align}
In periodic azimuth mode, the implementation uses analytical boxcar and Gaussian transfer functions; discrete boxcar equivalence is retained for the complete-data reflective/discrete formulation in Eq.~\ref{eq:tf_boxcar_discrete}.

Table~\ref{tab:frequency_vars} summarizes frequency-variable definitions.

\begin{table}[htbp]
	\centering
	\caption{Frequency variables for reflective and periodic transfer functions.}
	\label{tab:frequency_vars}
	\begin{tabular}{lccc}
		\toprule
		Boundary         & Boxcar frequency      & Gaussian/discrete frequency & Transform      \\
		\midrule
		Reflective (DCT) & $\theta_k=\pi k/(2N)$ & $\omega_k=\pi k/N$          & DCT-II (ortho) \\
		Periodic (RFFT)  & $\theta_k=\pi k/N$    & $\omega_k=2\pi k/N$         & RFFT (ortho)   \\
		\bottomrule
	\end{tabular}
\end{table}

\subsection{Cartesian Separability in D Dimensions}

For Cartesian $D$-dimensional smoothing,
\begin{equation}
	H_D[k_1,\ldots,k_D] = \prod_{d=1}^{D} H_d[k_d].
	\label{eq:nd_separable}
\end{equation}
This separability yields the axis-wise transfer-function multiplication path used in implementation, reducing the $D$-dimensional operation to sequential 1-D spectral multiplications while preserving spectral-efficiency advantages for large windows.

\section{Normalized Convolution Statistics}
\label{sec:nc_statistics}

This section defines missing-data-aware local statistics as normalized spectral operators \citep{Knutsson1993,Westin1994,Pham2006}, with practical mask-weighted context \citep{Astropy2022}.

\subsection{Mean, Moments, and Variance}

The local mean is
\begin{equation}
	\mu(\mathbf{i}) = \frac{\mathcal{S}_g\{x\cdot m\}(\mathbf{i})}{\mathcal{S}_g\{m\}(\mathbf{i})}.
	\label{eq:mean}
\end{equation}
For the reflective Cartesian operator chain,
\begin{equation}
	\mu(\mathbf{i}) = \frac{\mathrm{IDCT}\!\left\{\mathrm{DCT}\{x\cdot m\}\cdot H_g\right\}(\mathbf{i})}{\mathrm{IDCT}\!\left\{\mathrm{DCT}\{m\}\cdot H_g\right\}(\mathbf{i})}.
	\label{eq:mean_explicit}
\end{equation}

To avoid negative variance from floating-point cancellation in low-support regions, the variance is clamped before the square root:
\begin{align}
	E[X^2](\mathbf{i})   & = \frac{\mathcal{S}_g\{x^2\cdot m\}(\mathbf{i})}{\mathcal{S}_g\{m\}(\mathbf{i})}, \label{eq:second_moment} \\
	\sigma^2(\mathbf{i}) & = \max\!\left(0,\,E[X^2](\mathbf{i})-\mu(\mathbf{i})^2\right), \label{eq:var_clamped}                      \\
	\sigma(\mathbf{i})   & = \sqrt{\sigma^2(\mathbf{i})}. \label{eq:std}
\end{align}
This is population-style local variance within window support, not a Bessel-corrected sample variance.

\subsection{Effective Support and Stability Rule}

Effective local sample count (effective support) is
\begin{equation}
	N_{\mathrm{eff}}(\mathbf{i}) = \rho(\mathbf{i})\cdot A(\mathbf{i}), \qquad \rho(\mathbf{i})=\mathrm{clip}\!\left(\mathcal{S}_g\{m\}(\mathbf{i}),0,1\right).
	\label{eq:neff}
\end{equation}
where $A(\mathbf{i})$ is the local window area in grid units. For Cartesian $D$-dimensional windows, $A(\mathbf{i})=\prod_{d=1}^{D} W_d$ with axis-wise widths $W_d$. For polar smoothing, $A(\mathbf{i})$ is range-dependent and follows Eq.~\ref{eq:area_polar}.

The stable mean uses a denominator floor with fallback:
\begin{equation}
	\mu(\mathbf{i})=
	\begin{cases}
		\dfrac{\mathcal{S}_g\{x\cdot m\}(\mathbf{i})}{\mathcal{S}_g\{m\}(\mathbf{i})}, & \mathcal{S}_g\{m\}(\mathbf{i})\ge\epsilon, \\
		\mathcal{S}_g\{\bar{x}\}(\mathbf{i}),                                          & \mathcal{S}_g\{m\}(\mathbf{i})<\epsilon,
	\end{cases}
	\label{eq:mean_stable}
\end{equation}
where $\bar{x}$ is the prefilled field.
The prefilled field is produced by iterative spectral smoothing with nearest-neighbor completion, and this fallback replacement is applied only in under-supported regions.
The same denominator-floor rule applies to the second moment $E[X^2]$ in Eq.~\ref{eq:second_moment}, so the variance and standard deviation in Eqs.~\ref{eq:var_clamped}--\ref{eq:std} inherit the same stability.

Table~\ref{tab:nc_flow} summarizes the normalized-convolution algorithm flow.

\begin{table}[htbp]
	\centering
	\caption{Normalized-convolution algorithm flow for local statistics.}
	\label{tab:nc_flow}
	\begin{tabular}{ll}
		\toprule
		Step & Operation                                                              \\
		\midrule
		1    & Build mask $m$ and masked inputs ($x\cdot m$, $x^2\cdot m$)            \\
		2    & Smooth numerators and denominator with $\mathcal{S}_g$                 \\
		3    & Form normalized ratios for $\mu$ and $E[X^2]$                          \\
		4    & Compute $\sigma^2$ and $\sigma$ with clamp in Eq.~\ref{eq:var_clamped} \\
		5    & Apply denominator-floor rule in Eq.~\ref{eq:mean_stable}               \\
		6    & Use prefill and nearest completion only in under-supported regions     \\
		\bottomrule
	\end{tabular}
\end{table}

\section{Polar Smoothing with Adaptive Azimuth Kernels}
\label{sec:polar}

Polar arrays are indexed as $(n_{\mathrm{azimuth}}, n_{\mathrm{range}})$. To maintain approximately constant physical azimuth smoothing width, the adaptive azimuth kernel width in grid units is
\begin{equation}
	W_{\mathrm{az,grid}}(r) = \frac{W_{\mathrm{azimuth}}}{r\,\Delta\theta},
	\label{eq:polar_width}
\end{equation}
where $r$ is the 1-based range index and $\Delta\theta$ is azimuth spacing in radians \citep{Doviak1979,Bringi2001,Fabry2015}.

\subsection{Two-Stage Polar Operator}

Stage 1 (azimuth):
\begin{equation}
	Y_{\mathrm{az}} = \mathcal{T}_{\mathrm{az}}^{-1}\!\left\{\mathcal{T}_{\mathrm{az}}\{X\}\odot H_{\mathrm{az}}\right\},
	\label{eq:polar_stage1}
\end{equation}
with $\mathcal{T}_{\mathrm{az}}$ as DCT (reflective) or RFFT (periodic).
Because $W_{\mathrm{az,grid}}(r)$ varies with range, $H_{\mathrm{az}}$ is range-dependent and is evaluated as a 2-D transfer-function array across azimuth and range.

Because $H_{\mathrm{az}}$ is range-dependent, the operation is not represented as a single global 2-D spectral multiplication. Range smoothing is therefore applied sequentially to the azimuth-filtered intermediate field.
Stage 2 (range, reflective):
\begin{equation}
	Y = \mathrm{IDCT}_1\!\left\{\mathrm{DCT}_1\{Y_{\mathrm{az}}\}\odot H_{\mathrm{range}}\right\}.
	\label{eq:polar_stage2}
\end{equation}

For effective-support normalization (Eq.~\ref{eq:neff}),
\begin{equation}
	A_{\mathrm{polar}}(r) = W_{\mathrm{range}}\cdot W_{\mathrm{az,grid}}(r).
	\label{eq:area_polar}
\end{equation}

Table~\ref{tab:boundary_modes} lists boundary modes used in the polar formulation.

\begin{table}[htbp]
	\centering
	\caption{Boundary modes for polar smoothing.}
	\label{tab:boundary_modes}
	\begin{tabular}{llll}
		\toprule
		Axis    & Option     & Transform & Appropriate for                 \\
		\midrule
		Azimuth & Reflective & DCT-II    & Non-circular data               \\
		Azimuth & Periodic   & RFFT      & $0^\circ/360^\circ$ wrap-around \\
		Range   & Reflective & DCT-II    & Radar range (implemented mode)  \\
		\bottomrule
	\end{tabular}
\end{table}

Relative to \citet{Valdivia2026RadarDCT}, this section extends interpolation-first mean-only smoothing to normalized multi-statistics with explicit mixed-boundary disclosure.

\section{Validation Results}
\label{sec:validation}

All validation figures and metrics are generated with \texttt{dct\_toolkit} (\url{https://github.com/JValdivia23/dct-toolkit}) at \href{https://github.com/JValdivia23/dct-toolkit/commit/9d3b60ed408b5be819ae0eee0276c56840b712a7}{commit 9d3b60e}.

We evaluate three targeted scenarios aligned with the extension claims: a Cartesian boundary-condition check, a synthetic 3D outlier-identification test, and a real-radar polar demonstration.

\subsection{Cartesian Boundary-Condition Effects}

We generate a 1D non-periodic synthetic signal ($N=200$) with a monotonic trend and additive Gaussian noise to prevent boundary matching. A matched Gaussian kernel ($W=15$ grid points) is applied under reflective DCT and periodic FFT assumptions. Evaluation reports all-domain and edge-only RMSE; the edge subset is defined as the first and last $W$ samples to isolate wrap-around artifacts under periodic treatment.

\begin{figure}[htbp]
	\centering
	\includegraphics[width=0.95\linewidth]{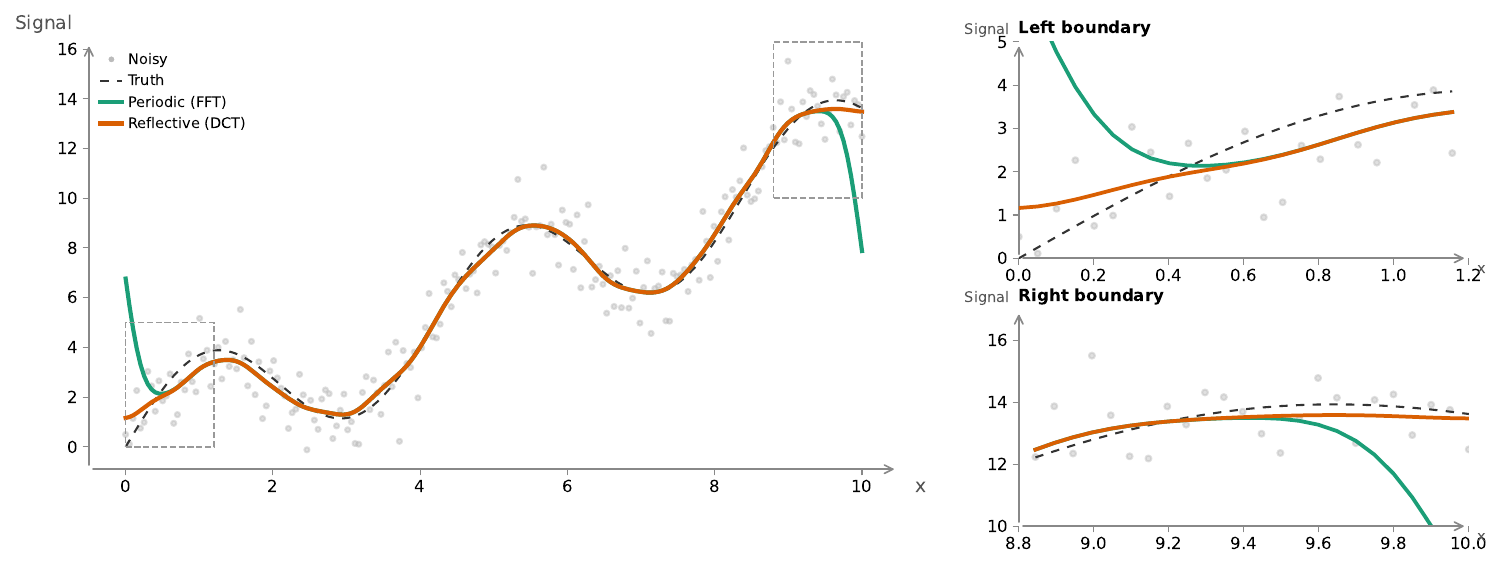}
	\caption{Boundary-condition sanity check on a 1D synthetic signal ($N=200$, $W=15$ Gaussian). Dashed boxes in the main panel mark edge regions expanded in the right-side zoom panels. Gray points denote noisy observations. Periodic FFT shows wrap-around artifacts (right edge pulled down, left edge pulled up), while reflective DCT follows local trend continuity. In this realization, reflective DCT shows a small left-right asymmetry: the left edge is slightly more positively biased than the right edge, consistent with different local slope configurations near each boundary.}
	\label{fig:boundary_sanity}
\end{figure}

Table~\ref{tab:boundary_results} and Figure~\ref{fig:boundary_sanity} show that reflective handling yields lower error than periodic handling in this 1D non-periodic setup. The dashed edge-region boxes and right-side zoom panels show where periodic wrap-around artifacts are strongest. The zoom panels also show a left-right asymmetry under reflective DCT in this realization: the right edge tracks the truth more closely, while the left edge shows a mild positive bias. This qualitative difference is consistent with local slope context near each boundary: at the right edge the local tendency bends downward relative to the global increasing background, while at the left edge both local and global tendencies increase, so reflective extension projects larger values into the averaging window.

\begin{table}[htbp]
	\centering
	\caption{Boundary-condition sanity check on a 1D non-periodic signal ($N=200$, $W=15$ Gaussian; 5 seeds, mean $\pm$ std).}
	\label{tab:boundary_results}
	\begin{tabular}{lrrr}
		\toprule
		Metric          & Reflective RMSE     & Periodic RMSE       & Relative increase (periodic) \\
		\midrule
		All-domain RMSE & $0.2798 \pm 0.0627$ & $1.0021 \pm 0.0400$ & 258.2\%                      \\
		Edge-only RMSE  & $0.3802 \pm 0.2030$ & $2.5161 \pm 0.0954$ & 561.8\%                      \\
		\bottomrule
	\end{tabular}
\end{table}

\subsection{3D Synthetic Outlier Identification}

We construct an idealized smooth 3D cyclone-like wind field with three vector components $(u,v,w)$ on an $18\times96\times96$ grid, add Gaussian perturbations ($\sigma=0.14$ per component), and inject sparse random vector anomalies at known indices as the ground-truth outlier mask. Injected anomalies occupy $0.3\%$ of grid points and use random 3D direction with amplitude sampled uniformly from $[1.25,2.25]$. This setup is intentionally challenging for global moments: signed wind components are approximately mean-centered at the domain scale, while physically valid cyclone structure produces large global spread. Outlier scoring is therefore performed on wind-speed residuals using local normalized-convolution statistics with Cartesian window widths $(W_z,W_y,W_x)=(3,7,7)$: $|\|\mathbf{v}\|-\mu_{\|\mathbf{v}\|}| > k\,\sigma_{\|\mathbf{v}\|}$ over $k\in\{2.0,2.5,\ldots,6.0\}$. Detection skill is summarized by precision (fraction of detected points that are true anomalies), recall (fraction of true anomalies that are detected), and the F1 score (their harmonic mean), consistent with local smoothing and window-tradeoff framing \citep{Savitzky1964,Harris1978}.
In operational terms, the local mean provides the expected neighborhood value, the local standard deviation characterizes the expected neighborhood spread, and a point is flagged when its local z-score---its deviation from the neighborhood mean in units of neighborhood standard deviation---exceeds a chosen threshold $k$.

\begin{figure}[htbp]
	\centering
	\includegraphics[width=0.88\linewidth]{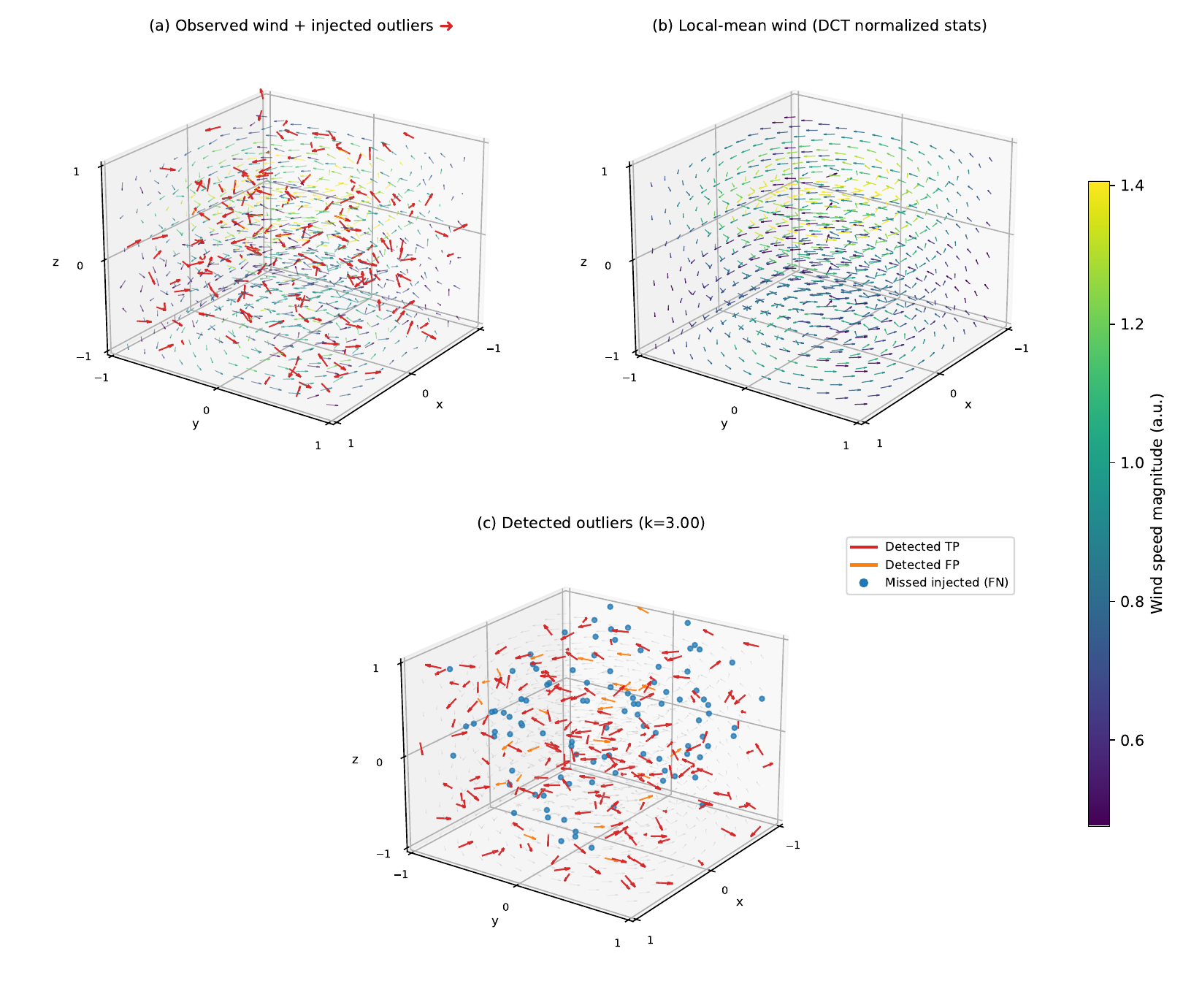}
	\caption{Synthetic 3D cyclone-wind outlier-identification example using sparse quiver rendering. (a)~Observed vector field with injected anomalies (red arrows). (b)~Local-mean vector field from normalized convolution. (c)~Detected outliers at $k=3.0$, with true positives (red), false positives (orange), and missed injected points (blue). Top row: panels~(a) and~(b); bottom row: panel~(c).}
	\label{fig:outlier_slices}
\end{figure}

\begin{figure}[htbp]
	\centering
	\includegraphics[width=0.88\linewidth]{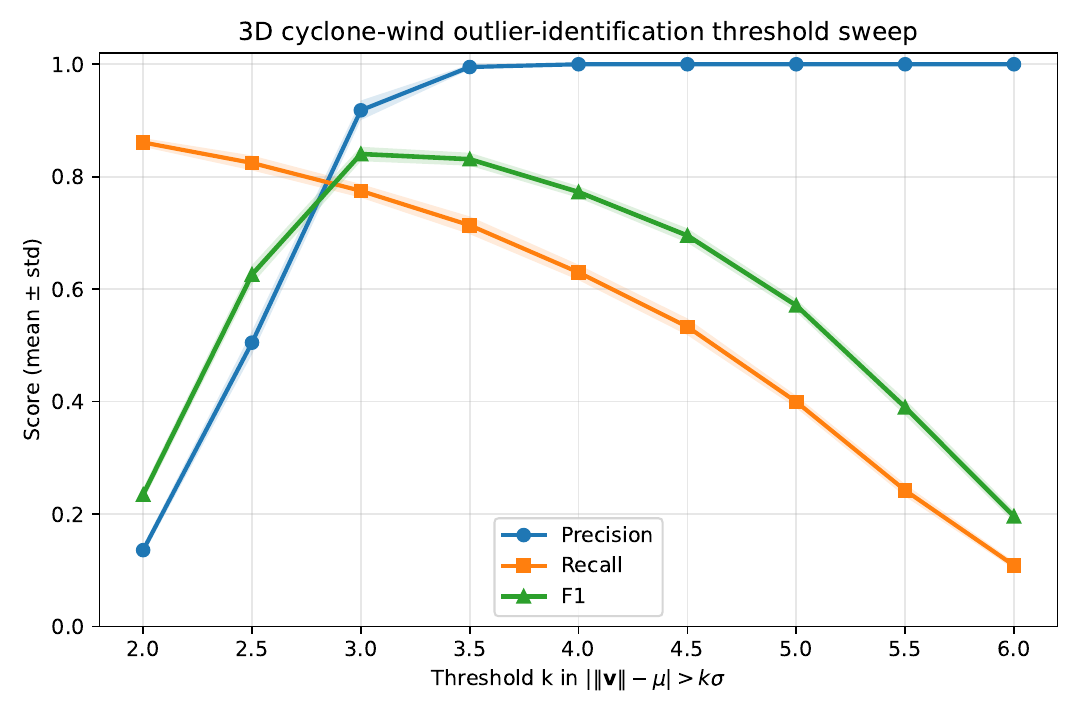}
	\caption{Threshold sweep for 3D cyclone-wind outlier identification using $|\|\mathbf{v}\|-\mu_{\|\mathbf{v}\|}| > k\sigma_{\|\mathbf{v}\|}$, reporting precision, recall, and F1 (mean $\pm$ std across seeds).}
	\label{fig:outlier_threshold}
\end{figure}

Table~\ref{tab:outlier_sweep_summary} and Figures~\ref{fig:outlier_slices}--\ref{fig:outlier_threshold} summarize threshold-sweep behavior for the cyclone-wind protocol. Figure~\ref{fig:outlier_slices}(a) shows that injected anomalies (red vectors) are sparse relative to the structured background flow, Figure~\ref{fig:outlier_slices}(b) shows the local-mean field estimated by normalized convolution, and Figure~\ref{fig:outlier_slices}(c) shows the detected mask at the selected threshold ($k=3.0$) with true positives, false positives, and missed injected points separated visually. Lower $k$ yields high recall but reduced precision, while larger $k$ produces near-perfect precision with increasing miss rate. Mean F1 peaks at $k=3.0$ ($0.8404\pm0.0129$), indicating a practical operating region that balances false alarms and missed detections under structured 3D flow. The result is consistent with the motivating failure mode of global moments in vortical flow: global mean-centering and large global spread are dominated by valid dynamics, whereas local normalization adapts to neighborhood context and preserves anomaly contrast.

\begin{table}[htbp]
	\centering
	\caption{Synthetic 3D cyclone-wind outlier-identification threshold sweep (5 seeds, mean $\pm$ std).}
	\label{tab:outlier_sweep_summary}
	\begin{tabular}{lccc}
		\toprule
		Threshold $k$ & Precision           & Recall              & F1                  \\
		\midrule
		2.0           & $0.1359 \pm 0.0077$ & $0.8609 \pm 0.0078$ & $0.2347 \pm 0.0116$ \\
		3.0           & $0.9181 \pm 0.0175$ & $0.7749 \pm 0.0120$ & $0.8404 \pm 0.0129$ \\
		4.0           & $1.0000 \pm 0.0000$ & $0.6298 \pm 0.0137$ & $0.7728 \pm 0.0103$ \\
		5.0           & $1.0000 \pm 0.0000$ & $0.3997 \pm 0.0108$ & $0.5711 \pm 0.0111$ \\
		5.5           & $1.0000 \pm 0.0000$ & $0.2423 \pm 0.0109$ & $0.3900 \pm 0.0141$ \\
		6.0           & $1.0000 \pm 0.0000$ & $0.1087 \pm 0.0066$ & $0.1960 \pm 0.0107$ \\
		\bottomrule
	\end{tabular}
\end{table}

\subsection{Polar Real-Radar Application}

Using the real-radar workflow from \texttt{dct\_toolkit/notebooks/01\_getting\_started.ipynb}, we apply two nominal smoothing-width settings ($W=5$ and $W=15$), where $W$ defines both the range width ($W_{\mathrm{range}}$) and azimuth target width ($W_{\mathrm{azimuth}}$) in Eq.~\ref{eq:polar_width}. The dataset is X-band radar (DOW6) data from the WINTRE-MIX field campaign (2022-02-23) and contains large missing sectors due to beam blockage.

\begin{figure}[htbp]
	\centering
	\includegraphics[width=0.95\linewidth]{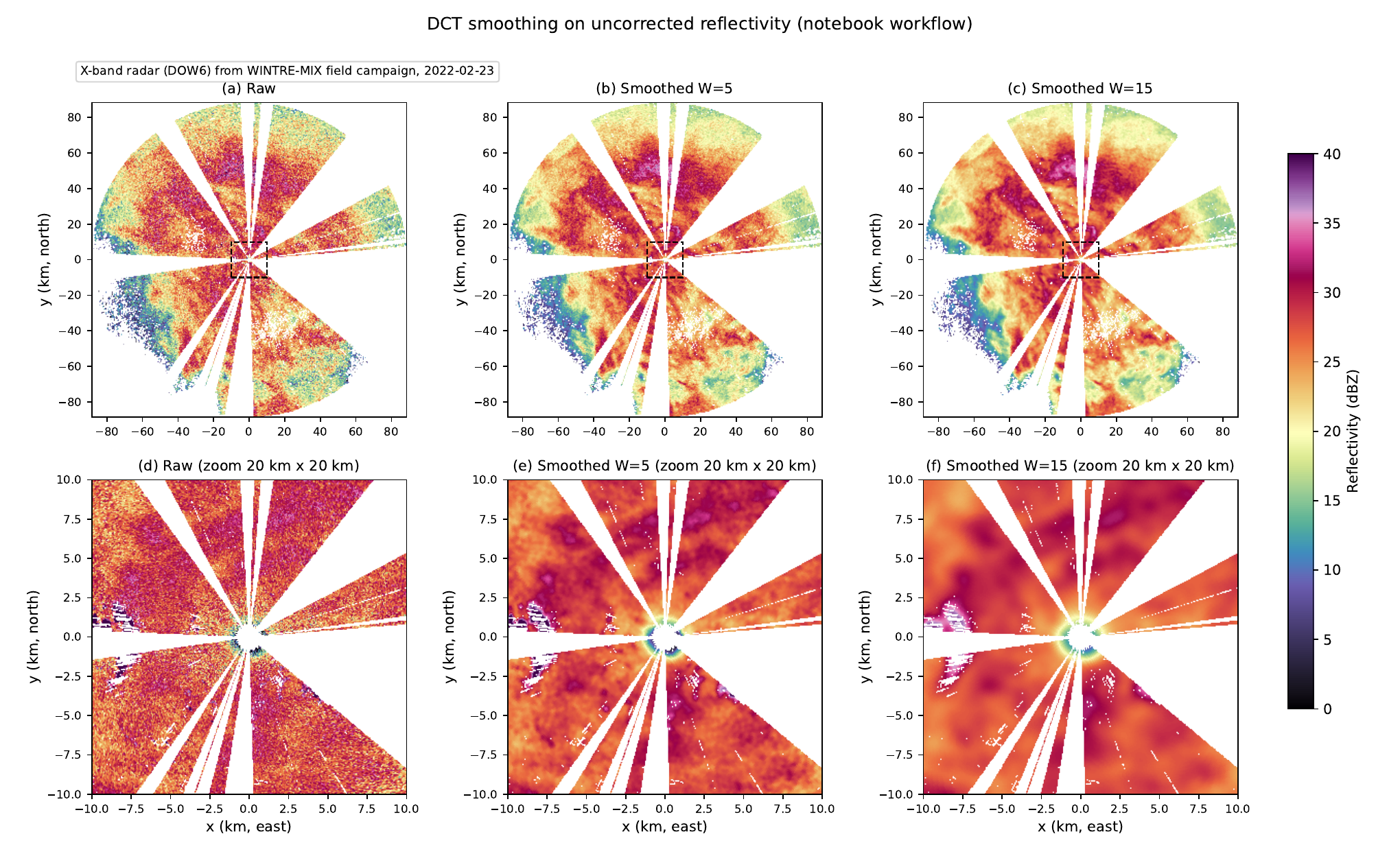}
	\caption{Real-radar polar demonstration on uncorrected DOW6 X-band reflectivity from WINTRE-MIX (2022-02-23), with shared color scale. Panels (a), (b), and (c) show the full-domain raw field, $W=5$ smoothing, and $W=15$ smoothing. Panels (d), (e), and (f) show the corresponding 20~km $\times$ 20~km central zoom. The comparison highlights mixed polar boundaries (periodic azimuth, reflective range), robust behavior under beam-blockage gaps, and no evident center distortion at $W=15$.}
	\label{fig:real_radar_6panel}
\end{figure}

Figure~\ref{fig:real_radar_6panel} shows a 6-panel comparison for the mixed-boundary polar operator. Panels (a)--(c) show full-domain views for raw reflectivity, $W=5$, and $W=15$, and panels (d)--(f) show the corresponding 20~km $\times$ 20~km central zoom. The comparison highlights three qualitative points. First, the combined periodic-azimuth and reflective-range treatment yields stable spatial behavior in the polar field. Second, even for the larger window ($W=15$), no evident center-of-grid distortion appears in panel (c) or panel (f). Third, despite extensive beam-blockage sectors in the raw field, normalized convolution still yields coherent local-mean estimates without obvious gap-driven artifacts in panels (b), (c), (e), and (f). As expected, $W=5$ preserves sharper local gradients while $W=15$ suppresses more small-scale variability, consistent with standard smoothing trade-offs \citep{Savitzky1964,Harris1978,Bringi2001,Fabry2015}. The evidence here is qualitative and dataset-specific.

\section{Discussion}
\label{sec:discussion}

Relative to \citet{Valdivia2026RadarDCT}, the present framework changes both the statistical target and missing-data treatment. The prior workflow centered on interpolation-first polar mean smoothing, whereas the current method formulates normalized convolution directly in the operators and extends outputs to variance, standard deviation, and effective support count. This is a methodological extension grounded in normalized-convolution theory \citep{Knutsson1993,Westin1994,Pham2006}.

The cyclone-wind validation highlights a key motivation for local statistics: in sign-changing rotating flow, global signed-component means are near zero while global variance remains large due to physically valid shear and curvature. Under this condition, global-threshold diagnostics are weakly discriminative for sparse anomalies. Local $\mu$ and $\sigma$ mitigate this by re-centering and re-scaling detection to neighborhood structure.

Section~\ref{sec:validation} supports a practical interpretation of boundary handling: boundary mode is an explicit model assumption. As shown by the 1D boundary sanity check, imposing periodic assumptions on non-periodic Cartesian signals introduces structural wrap-around artifacts that can propagate into local mean and variance interpretation near edges. Reflective boundaries are therefore appropriate for non-periodic Cartesian fields, while periodic azimuth handling is physically consistent for circular polar scans \citep{Strang1999,Ng1999,Martucci1994,Bringi2001,Fabry2015}.

The real-radar comparison illustrates the expected window-size trade-off: narrower windows preserve local detail but suppress variability less aggressively, while broader windows increase smoothing at the cost of fine-scale structure \citep{Savitzky1964,Harris1978}. This is a tuning choice, not evidence for a single globally optimal width.

Interpretation of Section~\ref{sec:validation} is intentionally bounded to extension-specific behavior (boundary effects, 3D outlier screening, and polar application). The experiments do not re-prove normalized convolution fundamentals \citep{Knutsson1993,Westin1994,Pham2006} and do not re-benchmark transform-scaling theory established in prior DCT work \citep{Valdivia2026RadarDCT}.

Several limitations remain. Outlier screening depends on threshold $k$, kernel width, local support ($N_{\mathrm{eff}}$), and boundary mode. Low-support regions can remain numerically stable but weakly constrained for inference. Operationally, users should mask or exclude estimates where $N_{\mathrm{eff}}$ falls below a chosen fraction of the nominal window area (e.g., $N_{\mathrm{eff}} < 0.5 \, A$), because variance and anomaly scores in under-supported regions are weakly constrained regardless of numerical stability. An iterative workflow is recommended: compute local mean and standard deviation, flag outliers using the z-score criterion, update the mask to exclude flagged points, and recompute the statistics to prevent outlier contamination from inflating local spread and masking genuine anomalies. The real-radar block is qualitative and dataset-specific, and broad cross-domain generalization requires additional benchmarks. Non-binary certainty-mask scenarios are not evaluated in this manuscript.

\section{Conclusion}
\label{sec:conclusion}

In summary, this work extends spectral smoothing into a practical local-statistics framework for incomplete gridded data by combining support-normalized estimators with explicit boundary-aware transforms. The validation results show that the correct boundary treatment matters for non-periodic Cartesian fields, that local mean and standard deviation can provide effective neighborhood-based anomaly screening in structured 3D flow, and that the mixed-boundary polar formulation behaves plausibly on real radar data with substantial missing sectors. These results suggest that the method is a useful option for radar and related geoscience workflows where fast, support-aware local statistics are needed on incomplete Cartesian or polar grids. Future work should test the framework more broadly against alternative baselines, expand sensitivity analyses, and evaluate non-binary certainty formulations.

\bibliographystyle{unsrtnat}
\bibliography{references}

\end{document}